\documentclass[twocolumn,aps]{revtex4-1}

\usepackage{xfrac}
\usepackage{amsmath}
\usepackage{amssymb}
\usepackage{xcolor}
\usepackage{hyperref}
\usepackage{graphics}

\newcommand{\T}[1]{\text{#1}}

\newcommand{\bra}[1]{\ensuremath{\langle #1 \vert}}
\newcommand{\ket}[1]{\ensuremath{\vert #1  \rangle}}
\newcommand{\braket}[2]{\ensuremath{\langle  #1 \vert #2  \rangle}}

\renewcommand\b[1]{\ensuremath{\mathbf{#1}}}

\begin{document}

\title{Fractional-charge and fractional-spin errors in range-separated density-functional theory}

\author{Bastien Mussard$^{a,b}$}\email{bastien.mussard@upmc.fr}
\author{Julien Toulouse$^{b}$}\email{julien.toulouse@upmc.fr}
\affiliation{
$^a$Institut des sciences du calcul et des donn\'ees, Universit\'e Pierre et Marie Curie, Sorbonne Universit\'es, F-75005, Paris, France\\
$^b$Laboratoire de Chimie Th\'eorique, Universit\'e Pierre et Marie Curie, Sorbonne Universit\'es, CNRS, F-75005, Paris, France}

\date{June 1, 2016}

\begin{abstract}
We investigate fractional-charge and fractional-spin errors in range-separated density-functional theory. Specifically, we consider the range-separated hybrid (RSH) method which combines long-range Hartree-Fock (HF) exchange with a short-range semilocal exchange-correlation density functional, and the RSH+MP2 method which adds long-range second-order M{\o}ller-Plesset (MP2) correlation. Results on atoms and molecules show that the fractional-charge errors obtained in RSH are much smaller than in the standard Kohn-Sham (KS) scheme applied with semilocal or hybrid approximations, and also generally smaller than in the standard HF method. The RSH+MP2 method tends to have smaller fractional-charge errors than standard MP2 for the most diffuse systems, but larger fractional-charge errors for the more compact systems. Even though the individual contributions to the fractional-spin errors in the H atom coming from the short-range exchange and correlation density-functional approximations are smaller than the corresponding contributions for the full-range exchange and correlation density-functional approximations, RSH gives fractional-spin errors that are larger than in the standard KS scheme and only slightly smaller than in standard HF. Adding long-range MP2 correlation only leads to infinite fractional-spin errors. This work clarifies the successes and limitations of range-separated density-functional theory approaches for eliminating self-interaction and static-correlation errors.\\
\end{abstract}

\maketitle

\section{Introduction}

The study of systems with fractional electron numbers proved to be extremely useful to diagnose the errors of computational electronic-structure methods, in particular methods based on density-functional theory (DFT)~\cite{CohMorYan-SCI-08,CohMorYan-CR-12}. Using the grand-canonical ensemble formalism, the exact ground-state energy $E^N$ of a system with a fractional electron number $N=M+\delta$, where $M$ is an integer and $0\le\delta\le 1$, was showed to be a linear function of $\delta$, interpolating between the ground-state energies $E^M$ and $E^{M+1}$ of the adjacent $M$- and $(M+1)$-electron systems~\cite{PerParLevBal-PRL-82} (see also Ref.~\onlinecite{YanZhaAye-PRL-00} for a pure-state view)
\begin{equation}
  E^N=(1-\delta)E^M+\delta E^{M+1}.
\label{EN}
\end{equation}
The deviation from this piecewise linearity behavior of the energy obtained with a given approximate method is a measure of the (many-electron) self-interaction error~\cite{MorCohYan-JCP-06,RuzPerCsoVydScu-JCP-06,RuzPerCsoVydScu-JCP-07,VydScuPer-JCP-07}, also called the delocalization error or the fractional-charge error~\cite{MorCohYan-PRL-08}, of this method. 

For systems with degenerate ground states with different spin, another useful condition can be derived. For example, for a system with a spin-doublet ground state, such as the hydrogen atom or the lithium atom, the exact ground-state energy $E^{N_\uparrow,N_\downarrow}$ with fractional spin-up and spin-down electron numbers, $N_\uparrow=M/2+1-\delta$ and $N_\downarrow=M/2+\delta$ where $M$ is an even integer and $0\le\delta\le 1$, is independent of $\delta$
\begin{equation}
E^{N_\uparrow,N_\downarrow}=E^{M/2+1,M/2}=E^{M/2,M/2+1}.
\label{ENupdn}
\end{equation}
This is known as the constancy condition~\cite{CohMorYan-JCP-08}. The deviation from this constancy condition of the energy obtained with a given approximate method is a measure of the static (or strong) correlation error, also called fractional-spin error in this context. The conditions of Eqs.~(\ref{EN}) and~(\ref{ENupdn}) can be unified and extended to the so-called flat-plane condition~\cite{MorCohYan-PRL-09}. 

An ideal computational electronic-structure method should fulfill the two conditions of Eqs.~(\ref{EN}) and~(\ref{ENupdn}). Let us first focus on the deviation from Eq.~(\ref{EN}), i.e. on the fractional-charge error.
Semilocal density-functional approximations (DFAs), i.e. the local-density approximation (LDA), generalized-gradient approximations (GGA), and meta-GGA approximations, tend to give convex energy curves as a function of $N$, i.e. favoring too much electron delocalization~\cite{MorCohYan-JCP-06,VydScuPer-JCP-07}. On the opposite, Hartree-Fock (HF) gives concave energy curves, i.e. favoring too much electron localization~\cite{MorCohYan-JCP-06,VydScuPer-JCP-07}. Hybrid approximations combining a fraction of HF exchange with a semilocal DFA help to reduce fractional-charge errors, even though they still give significantly convex energy curves~\cite{MorCohYan-JCP-06,VydScuPer-JCP-07}. Long-range corrected (LC)~\cite{IikTsuYanHir-JCP-01} or Coulomb-attenuated-method (CAM)~\cite{YanTewHan-CPL-04} approximations introducing 100\% of HF exchange at long-range electron-electron distances tend to further reduce fractional-charge errors~\cite{MorCohYan-JCP-06,VydScuPer-JCP-07,CohMorYan-JCP-07,TsuSonSuzHir-JCP-10,MalPenYanBalHol-TCA-14}. Moving now to approximations depending on virtual orbitals, second-order M{\o}ller-Plesset (MP2) perturbation theory was found to give very small fractional-charge errors~\cite{CohMorYan-JCTC-09,SuXu-JCTC-15}. Double-hybrid (DH) approximations~\cite{Gri-JCP-06}, combining HF exchange and MP2 correlation with a semilocal DFA, inherit from the good fractional-charge behavior of standard MP2, and lead to smaller fractional-charge errors than standard hybrid approximations, even though the errors are generally larger than in standard MP2~\cite{SuYanMorXu-JPCA-14,SuXu-MP-16}. The random-phase approximation (RPA) in its simplest direct variant~\cite{Fur-PRB-01} gives quite convex energy curves as a function of $N$, corresponding to a large delocalization error~\cite{MorCohYan-PRA-12,HelRohGro-JCP-12,HelCarRohRenRubSchRin-PRB-15}. This fractional-charge error can be very much reduced by adding the HF exchange kernel within the standard (particle-hole) RPA approach~\cite{MorCohYan-PRA-12} or within the particle-particle RPA approach~\cite{AggYanYan-PRA-13}. Finally, coupled-cluster methods with single and double excitations (CCSD) and perturbative triple excitations [CCSD(T)] have also been extended to fractional electron numbers and have been found to nearly satisfy the exact piecewise linear behavior~\cite{SteYan-JCP-13}.

Let us now focus on the deviation from Eq.~(\ref{ENupdn}), i.e. on the fractional-spin error. It turns out that most of the symmetry-unbroken single-determinant-reference methods giving small fractional-charge errors tend to give large fractional-spin errors, and vice versa. For example, while the addition of HF exchange in hybrid approximations helps to decrease fractional-charge errors, as previously mentioned, it also increases fractional-spin errors in comparison to semilocal DFAs~\cite{CohMorYan-JCP-08}. Even more illustrative is the case of MP2 which is among the methods giving the smallest fractional-charge errors, and also gives infinite fractional-spin errors~\cite{CohMorYan-JCTC-09}. This last problem is partly overcome by self-consistent MP2 based on Green-function theory which gives finite (but not very small) fractional-spin errors in the fractional-spin H atom~\cite{PhiKanZgi-JCP-15}. Direct RPA has no fractional-spin errors for the fractional-spin H atom~\cite{MorCohYan-PRA-12,HelRohGro-JCP-12}, while CCSD shows fractional-spin errors in systems with more than two degenerate fractionally occupied orbitals. So far, only particle-particle RPA seems to be able to give both small fractional-charge and fractional-spin errors~\cite{AggYanYan-PRA-13}.

In this work, we investigate the fractional-charge and fractional-spin errors in range-separated DFT (see, e.g., Refs.~\onlinecite{Sav-INC-96,TouColSav-PRA-04}) for some atoms (H, He, Li, Be, B, C, N, O, F) and molecules (N$_2$ and CO). We first focus on the range-separated hybrid (RSH) approximation which combines long-range HF exchange with a short-range exchange-correlation DFA~\cite{PedJen-JJJ-XX,AngGerSavTou-PRA-05,FroTouJen-JCP-07}, which is similar but not identical to the LC scheme. In particular, we analyze the contributions to the fractional-charge and fractional-spin errors coming from the short-range exchange and correlation DFAs. We then also study the effect of adding a long-range correlation energy calculated at the MP2 level~\cite{AngGerSavTou-PRA-05,FroJen-PRA-08,Ang-PRA-08}. Our work helps to clarify the successes and limitations of range-separated DFT approaches for eliminating self-interaction and static-correlation errors.

\section{Range-separated DFT with fractional electron numbers}

\subsection{Range-separated hybrid approximation}

The extension of the RSH scheme~\cite{AngGerSavTou-PRA-05} to fractional electron numbers is easily performed, just like for HF or other hybrid Kohn-Sham approximations~\cite{MorCohYan-JCP-06}, by introducing fractional occupation numbers $n_k$ for the RSH orbitals $\varphi_k(\b{r})$. The total electronic energy in the spin-unrestricted RSH approximation is thus written as, in atomic units,
\begin{eqnarray}
E_\T{RSH} &=& \sum_k n_k \int \varphi_k^*(\b{r}) \left[ -\frac{1}{2}\nabla^2 + v_\T{ne}(\b{r}) \right] \varphi_k(\b{r})  \T{d}\b{r} 
\nonumber\\
&& + E_\T{H}[n] + E_\T{x}^\T{lr} + E^\T{sr}_\T{xc}[n_\uparrow,n_\downarrow],
\end{eqnarray}
where $k$ is a spin-orbital index implicitly including a spin index ($\sigma=\uparrow,\downarrow$), $v_\T{ne}(\b{r})$ is the nuclei-electron interaction, $E_\T{H}[n]$ is the full-range Hartree energy depending on the total density $n(\b{r})=\sum_\sigma n_\sigma(\b{r})$
\begin{eqnarray}
E_\T{H}[n]= \frac{1}{2} \iint n(\b{r}_1) n(\b{r}_2) w_\T{ee}(|\b{r}_2-\b{r}_1|) \T{d} \b{r}_1 \T{d} \b{r}_2,
\label{EH}
\end{eqnarray}
with the standard Coulomb electron-electron interaction $w_\T{ee}(r)=1/r$, $E_\T{x}^\T{lr}$ is the long-range HF exchange energy depending on the spin-resolved one-particle density matrix $\gamma_\sigma(\b{r}_1,\b{r}_2)$
\begin{eqnarray}
E_\T{x}^\T{lr}= -\frac{1}{2} \sum_\sigma \iint |\gamma_\sigma(\b{r}_1,\b{r}_2)|^2 w_\T{ee}^\T{lr}(|\b{r}_2-\b{r}_1|) \T{d} \b{r}_1 \T{d} \b{r}_2, \;\;
\label{Exlr}
\end{eqnarray}
with the long-range electron-electron interaction $w_\T{ee}^\T{lr}(r)=\T{erf}(\mu r)/r$ and the range-separation parameter $\mu$, and $E^\T{sr}_\T{xc}[n_\uparrow,n_\downarrow]$ is the short-range exchange-correlation energy depending on the spin densities $n_\sigma(\b{r})$. The spin densities and the one-particle density matrix are expressed with the fractional orbital occupation numbers
\begin{eqnarray}
n_{\sigma}(\b{r}) = \sum_{k \text{ of spin } \sigma} n_k |\varphi_k(\b{r})|^2,
\label{n}
\end{eqnarray}
and
\begin{eqnarray}
\gamma_{\sigma}(\b{r},\b{r}') = \sum_{k \text{ of spin } \sigma} n_k \varphi_k^*(\b{r}') \varphi_k(\b{r}),
\label{gamma}
\end{eqnarray}
where the sum is over all spin orbitals $k$ of spin $\sigma$.

For $\mu=0$, fractional occupation number RSH reduces to Kohn-Sham DFT, while for $\mu\to\infty$ it reduces to the HF method, both in their extended versions to fractional occupation numbers. The RSH approximation is similar to the LC scheme~\cite{IikTsuYanHir-JCP-01}, except that in the latter a full-range correlation functional is used instead of a short-range correlation functional. We note that other definitions of the Hartree and exact exchange contributions for an ensemble than the ones of Eqs.~(\ref{EH}) and~(\ref{Exlr}) are possible (see, e.g., Refs.~\onlinecite{GouDob-JCP-13,SenHedAlaKneFro-MP-16}). The ones of Eqs.~(\ref{EH}) and~(\ref{Exlr}), based on the ensemble density of Eq.~(\ref{n}) and the ensemble one-particle density matrix of Eq.~(\ref{gamma}), are the relevant ones for the purpose of analyzing fractional systems arising for example from molecular dissociations.

In practice, in an atomic-orbital (AO) basis $\{ \chi_\mu(\b{r}) \}$, the total RSH electronic energy is calculated as
\begin{eqnarray}
E_\T{RSH} = \sum_{\mu \nu} P_{\nu \mu} \left[ h_{\mu \nu} + \frac{1}{2} \left( J_{\mu \nu} - K^\T{lr}_{\mu \nu} \right) \right] + E^\T{sr}_\T{xc}[n_\uparrow,n_\downarrow],
\nonumber\\
\end{eqnarray}
where $P_{\mu \nu} = \sum_\sigma P_{\mu \nu}^{\sigma}$ is the total AO density matrix, $h_{\mu \nu}$ are the one-electron (kinetic + electron-nuclei) integrals, $J_{\mu \nu} = \sum_{\lambda \gamma} P_{\gamma \lambda} \braket{\mu \lambda}{\nu \gamma}$ is the Hartree contribution, $K^\T{lr}_{\mu \nu} = \sum_{\lambda \gamma} P_{\gamma \lambda} \braket{\mu \lambda}{\gamma \nu}^\T{lr}$ is the long-range HF exchange contribution, and the spin densities are calculated as $n_\sigma(\b{r}) =
\sum_{\mu \nu} P_{\mu \nu}^{\sigma} \chi_\mu(\b{r}) \chi_\nu^*(\b{r})$ with the spin-resolved AO density matrix $P_{\mu \nu}^\sigma$. In these expressions, $\braket{\mu \lambda}{\nu \gamma}$ are the two-electron integrals for the Coulomb interaction $w_\T{ee}(r)$ and $\braket{\mu \lambda}{\gamma \nu}^\T{lr}$ are the two-electron integrals for the long-range interaction $w_\T{ee}^\T{lr}(r)$. The fractional occupation numbers $n_k$ only appear in the AO density matrix whose expression is
(assuming from now on real-valued orbitals),
\begin{eqnarray}
P_{\mu\nu}^\sigma =\sum_{k \text{ of spin } \sigma} n_{k} c_{k\mu} c_{k\nu},
\label{Pmunu}
\end{eqnarray}
where $c_{k\mu}$ is the coefficient of the spin orbital $k$ over the basis function $\chi_\mu$. 

Three sets of spin orbitals can be defined: $N_\T{f}$ \textit{fully occupied} spin orbitals for which $n_k=1$, $N_\T{p}$ \textit{partially occupied} spin orbitals for which $0 \leq n_k \leq 1$ (corresponding to possibly degenerate HOMO spin orbitals for each spin), and $N_\T{u}$ \textit{fully unoccupied} spin orbitals for which $n_k=0$. We always consider fixed occupation numbers $n_k$ which add up to the desired spin-up and spin-down fractional electron numbers $N_\uparrow$ and $N_\downarrow$. The RSH total energy is then minimized with respect to the orbital coefficients.

We note that, while for the standard case of non-fractional occupation numbers the non-redundant orbital-rotation parameters to optimize corresponds to \textit{occupied} $\to$ \textit{unoccupied} excitations, in the present case of fractional occupation numbers there are \textit{a priori} more orbital-rotation parameters to consider: \textit{fully occupied} $\to$ \textit{partially occupied}, \textit{fully occupied} $\to$ \textit{fully unoccupied}, \textit{partially occupied} $\to$ \textit{partially occupied}, and \textit{partially occupied} $\to$ \textit{fully unoccupied} (see, e.g., Ref.~\onlinecite{CanMou-NL-14}). Hence, the number of parameters is in principle $(N_\T{f} + N_\T{p})(N_\T{p} + N_\T{u})$. However, in practice, if using canonical spin orbitals diagonalizing
the Fock matrix, it is sufficient in a program to change the definition of the density matrix according to Eq.~(\ref{Pmunu}) and perform the orbital optimization as in the standard case where the partially occupied spin orbitals would be fully occupied, i.e. only optimizing $(N_\T{f} + N_\T{p}) N_\T{u}$ parameters. Indeed, the energy gradient with respect to the additional parameters only involve off-diagonal elements of the Fock matrix and are thus automatically zero with canonical spin orbitals.

\subsection{Long-range MP2 correlation energy}

In the RSH total energy, the long-range correlation energy $E_\T{c}^\T{lr}$ is missing and can be calculated \textit{a posteriori}, e.g. with long-range MP2~\cite{AngGerSavTou-PRA-05,FroJen-PRA-08,Ang-PRA-08}. The expression of the MP2 correlation energy extended to fractional electron numbers was given by Casida~\cite{Cas-PRB-99} and Cohen {\it et al.}~\cite{CohMorYan-JCTC-09}. The extension to the range-separated case is straightforward and is described now.

We calculate the long-range MP2 correlation energy as
\begin{equation}
E_\T{c}^\T{lr,MP2} = \frac{1}{2}\sum_{ia,jb} B^\T{x}_{ia,jb} T^\T{d,(1)}_{jb,ia},
\label{EclrMP2}
\end{equation}
where $B^\T{x}_{ia,jb} = f_{ia,jb} \bra{ab}\ket{ij}^\T{lr}$ are the exchange interaction matrix elements and $T^\T{d,(1)}_{jb,ia}$ are the direct first-order double-excitation amplitudes
\begin{equation}
T^{\T{d},(1)}_{ia,jb}=\frac{-B^\T{d}_{ia,jb}}{\varepsilon_a+\varepsilon_b-\varepsilon_i-\varepsilon_j},
\label{Td1}
\end{equation}
where $B^\T{d}_{ia,jb} = f_{ia,jb} \braket{ab}{ij}^\T{lr}$ are the direct interaction matrix elements. In these expressions, $\bra{ab}\ket{ij}^\T{lr}=\braket{ab}{ij}^\T{lr}-\braket{ab}{ji}^\T{lr}$ and $\braket{ab}{ij}^\T{lr}$ are the antisymmetrized and non-antisymmetrized long-range two-electron integrals over RSH spin orbitals, respectively, $\varepsilon_{k}$ are RSH spin-orbital energies, and $f_{ia,jb}=\sqrt{n_i n_j(1-n_a)(1-n_b)}$ is the fractional-occupation-number factor of
Ref.~\onlinecite{YanMorCoh-JCP-13}.

In Eq.~(\ref{EclrMP2}), the sums over $i$ and $j$ are over all fully occupied and partially occupied spin orbitals, while the sums over $a$ and $b$ are over all partially occupied and fully unoccupied spin orbitals. Note that, in practice, we restrict these sums to only non-spin-flip excitations $i\to a$ and $j\to b$ (i.e., the spin orbitals $i$ and $a$ have the same spins, and the spin orbitals $j$ and $b$ have the same spins) since $B^\T{d}_{ia,jb} = 0$ for spin-flip excitations. Also, we exclude from the sums of Eq.~(\ref{EclrMP2}) all terms for which $B^\T{x}_{ia,jb} = 0$, which include in particular the terms with $i=j$ or $a=b$.

For the range-separation parameter $\mu=0$, the long-range MP2 correlation energy of Eq.~(\ref{EclrMP2}) vanishes, while for $\mu\to\infty$ it reduces to the post-HF full-range MP2 correlation energy with fractional occupation numbers of Ref.~\onlinecite{CohMorYan-JCTC-09}.

\subsection{Computational details}

All calculations were done with a development version of {\tt MOLPRO 2015}~\cite{Molproshort-PROG-15}, in which we have implemented the above-described extensions to fractional occupation numbers. 

The RSH calculations are done with the short-range exchange-correlation Perdew-Burke-Ernzerhof (PBE) density functional of Ref.~\onlinecite{GolWerStoLeiGorSav-CP-06}, which is a modified version of the short-range functional of Ref.~\onlinecite{TouColSav-JCP-05} so that it reduces to the standard PBE functional~\cite{PerBurErn-PRL-96} for $\mu=0$. The range-separated MP2 total energy is obtained as
\begin{equation}
E_\T{RSH+MP2}=E_\T{RSH}+E_\T{c}^\T{lr,MP2},
\label{ERSH+corr}
\end{equation}
and is referred to as RSH+MP2. In these calculations, the range-separation parameter is fixed to $\mu=0.5$ bohr$^{-1}$, as recommended in previous studies~\cite{GerAng-CPL-05a,MusReiAngTou-JCP-15}.

For comparison, we also perform HF calculations and Kohn-Sham calculations using the PBE functional and the PBE0 hybrid approximation~\cite{AdaBar-JCP-99,ErnScu-JCP-99a}, as well as full-range MP2 calculations starting from HF, i.e.
\begin{equation}
E_\T{HF+MP2}=E_\T{HF}+E_\T{c}^\T{MP2},
\end{equation}
corresponding to Eq.~(\ref{ERSH+corr}) in the limit $\mu\to\infty$, which is referred to as HF+MP2. 

The calculations for H, He, H$_2^+$, He$_2^+$, and H$_2$ were done with the Dunning cc-pVTZ basis sets~\cite{Dun-JCP-89,WooDun-JCP-94}. The calculations for Li, Be, B, C, N, O, and F were done with the aug-cc-pCVTZ basis sets~\cite{WooDun-JCP-95}, including core excitations in the MP2 calculations. The calculations for N$_2$ and CO were done with the aug-cc-pVTZ basis sets~\cite{KenDunHar-JCP-92}, without core excitations. We note that using basis sets augmented with diffuse basis functions has an important impact on the fractional-charge errors for the negatively charged systems considered (see, in particular, Ref.~\onlinecite{PeaTeaHelToz-JCTC-15}). The internuclear distances of N$_2$ and CO were taken as the experimental distances of 1.098 \r{A}~\cite{HubHer-BOOK-79} and 1.128 \r{A}~\cite{NIST-BOOK-15}, respectively. Spatial symmetry is not explicitly imposed in our calculations (but not necessarily broken either). The calculations with fractional occupation numbers are always done in a spin-unrestricted formalism. For fractional-charge calculations, either the spin-up or spin-down HOMO orbital is fractionally occupied so as to connect the lowest-energy states (according to Hund's rule of maximum spin multiplicity) at both adjacent integer-electron numbers.

\section{Results and discussion}

In all figures, the SCF and MP2 results are color- and symbol-coded in a consistent way; full-range calculations are always shown with dotted lines while range-separated calculations are shown with full lines.

\subsection{Dissociation of H$_2^+$ and fractional-charge error in H}

\begin{figure*}
\includegraphics[width=.39\linewidth,clip=true,trim=0cm 0mm 0cm 0cm]{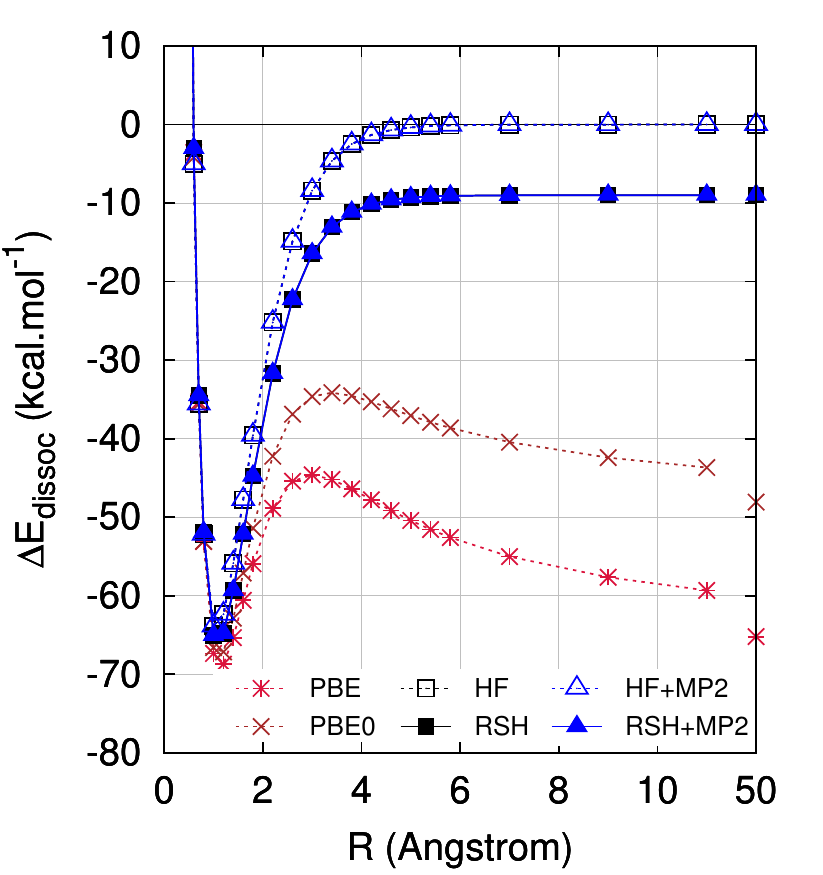}
\includegraphics[width=.39\linewidth,clip=true,trim=0cm 0mm 0cm 0cm]{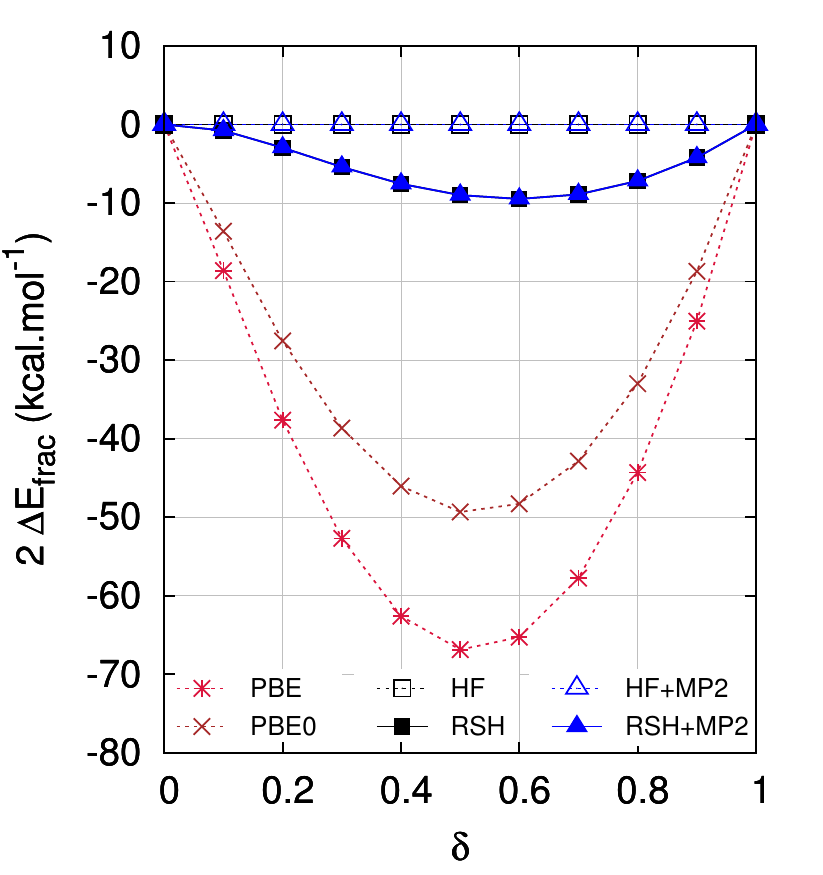}
\caption{Dissociation of H$_2^+$ and corresponding fractional-charge analysis for the H atom. On the left panel, the dissociation energy curve $\Delta E_\T{dissoc}(R)$ defined in Eq.~(\ref{h2+Edissoc}) is plotted as a function of the internuclear distance $R$ for each method. On the right panel, the fractional-charge error $2\; \Delta E_\T{frac}(\delta)$ defined in Eq.~(\ref{h2+Efrac}) is plotted as a function of $\delta$ for the same methods.}
\label{fig:h2+}
\end{figure*}

The study of the dissociation of H$_2^+$ reveals the propensity of each computational method to the one-electron self-interaction error. In the left panel of Figure~\ref{fig:h2+}, the total energy of H$_2^+$ calculated by each method, without breaking spatial symmetry in the calculation, is plotted as a function of the internuclear distance $R$, taking as zero energy reference the sum of the energies of the isolated ion $\T{H}^+$ and atom $\T{H}$ (with integer numbers of electrons) calculated with the same method
\begin{equation}
\Delta E_\T{dissoc}(R) = E(\T{H}_2^+,R) - \left[ E(\T{H}^+) + E(\T{H}) \right].
\label{h2+Edissoc}
\end{equation}
On the right panel of Figure~\ref{fig:h2+}, for each corresponding method, we plot twice the error in the total energy of a fractionally charged hydrogen atom, $\T{H}^{+(1-\delta)}$, with respect to the linear interpolation of the energies of $\T{H}^+$ and $\T{H}$ as a function of the fractional number of electron $\delta$ with $0\leq\delta\leq 1$
\begin{equation}
2\Delta E_\T{frac}(\delta) = 2 E(\T{H}^{+(1-\delta)}) - 2 \left[ (1-\delta) E(\T{H}^+) +\delta E(\T{H}) \right].
\label{h2+Efrac}
\end{equation}
In the exact case, $\Delta E_\T{frac}(\delta) = 0$ for any $0\leq\delta\leq 1$, according to Eq.~(\ref{EN}). For all the SCF methods considered here, H$_2^+$ dissociates into two half-electron hydrogen ions, i.e. $E(\T{H}_2^+,R\to\infty) = E(\T{H}^{+0.5} + \T{H}^{+0.5})$. Since for all the SCF and MP2 methods considered here the total energy is additively separable, i.e. $E(\T{H}^{+0.5} + \T{H}^{+0.5}) = 2 E(\T{H}^{+0.5})$, the dissociation limit of the energy curve is
\begin{eqnarray}
  \label{eq:h2+dissocfrac}
\Delta E_\T{dissoc}(R\to\infty) &=& 2 E(\T{H}^{+0.5}) - \left[ E(\T{H}^+) + E(\T{H}) \right]
\nonumber\\
&=&2 \Delta E_\T{frac}(\delta=0.5).
\end{eqnarray}
Thus, the value $2\; \Delta E_\T{frac}(\delta=0.5)$ corresponds to the one-electron self-interaction error in H$_2^+$ in the dissociation limit. More generally, the curve of the fractional-charge error $\Delta E_\T{frac}(\delta)$ is a convenient way to analyze the self-interaction error of a method. 
The equality of Eq.~(\ref{eq:h2+dissocfrac}) can be checked by comparing the left and right panels of Figure~\ref{fig:h2+} 
(note that the point displayed on the right of all dissociation figures is at 50 \AA).

Clearly, HF is exact for H$_2^+$ (it is one-electron self-interaction free) and correctly gives $\Delta E_\T{frac}(\delta)=0$ for all $0\leq\delta\leq 1$. By contrast, PBE leads to large fractional-charge errors. Due to the introduction of 25\% of global HF exchange, PBE0 has a bit smaller fractional-charge errors. RSH has much smaller fractional-charge errors than PBE0, which is what was also previously found for LC-$\omega$PBE~\cite{VydScuPer-JCP-07}. Note that both the PBE and PBE0 dissociation energy curves display a spurious barrier due to the electrostatic repulsion (proportional to $1/R$) between the charged fragments $\T{H}^{+0.5}$ and $\T{H}^{+0.5}$, as explained in Ref.~\onlinecite{GraKraCre-JCP-04}. This spurious barrier is absent with RSH thanks to the introduction of the long-range HF exchange which correctly cancels out the electrostatic repulsion of the charged fragments.

For this system, the MP2 correlation energy is correctly zero for any $0\leq\delta\leq 1$. Indeed, it can be easily checked that when a single spin orbital $i$ is occupied $B^\T{x}_{ia,ib}=0$ in Eq.~(\ref{EclrMP2}). Therefore, the HF+MP2 results are identical to the HF results, and the RSH+MP2 results are identical to the RSH results. The small fractional-charge errors of RSH+MP2 are thus only due to the use of the approximate short-range PBE exchange-correlation density functional.

\begin{figure}
\includegraphics[width=0.9\linewidth,clip=true,trim=0cm 0mm 0cm 0cm]{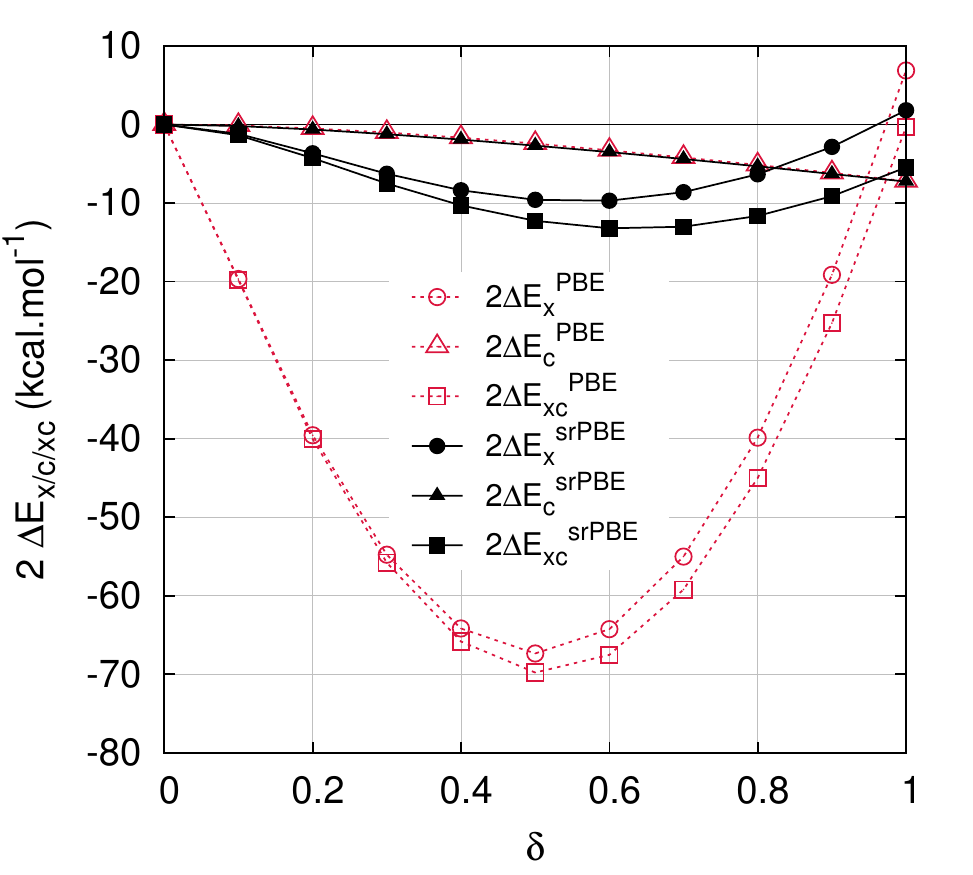}
\caption{Density-functional-approximation contributions to the fractional-charge error in the H atom as a function of the fractional electron number $\delta$. For Kohn-Sham PBE calculations, we show the error due to the PBE exchange energy, $\Delta E_\T{x}^\T{PBE} = E_\T{x}^\T{PBE}[n_\delta,0] + E_\T{H}[n_\delta]$, the PBE correlation energy $\Delta E_\T{c}^\T{PBE} = E_\T{c}^\T{PBE}[n_\delta,0]$, and their sum $\Delta E_\T{xc}^\T{PBE}=\Delta E_\T{x}^\T{PBE}+\Delta E_\T{c}^\T{PBE}$, where the density $n_\delta$ is from the self-consistent PBE calculation at the value of $\delta$. For the RSH calculations, we show the error due to the short-range PBE exchange energy, $\Delta E_\T{x}^\T{srPBE} = E_\T{x}^\T{srPBE}[n_\delta,0] + E_\T{H}^\T{sr}[n_\delta]$, the short-range PBE correlation energy $\Delta E_\T{c}^\T{srPBE} = E_\T{c}^\T{srPBE}[n_\delta,0]$, and their sum $\Delta E_\T{xc}^\T{srPBE} = \Delta E_\T{x}^\T{srPBE} + \Delta E_\T{c}^\T{srPBE}$, where the density $n_\delta$ is from the self-consistent RSH calculation at the value of $\delta$. The errors are multiplied by a factor of 2 to be directly comparable to Figure~\ref{fig:h2+}.
}
\label{fig:h_exc_frac}
\end{figure}

We can analyze further the contributions to the fractional-charge error coming from the approximate exchange and correlation density functionals used. For Kohn-Sham DFT applied to systems with one or less spin-up electron (see, e.g., Refs.~\onlinecite{PerZun-PRB-81,ZhaYan-JCP-98}), the exact exchange energy functional cancels out the Hartree energy
\begin{equation}
E_\T{x}[n_\delta,0] = - E_\T{H}[n_\delta],
\end{equation}
where the exact exchange functional is defined as $E_\T{x}[n_\uparrow,n_\downarrow]= -(1/2) \sum_\sigma \iint |\gamma_\sigma(\b{r}_1,\b{r}_2)|^2 w_\T{ee}(|\b{r}_2-\b{r}_1|) \T{d} \b{r}_1 \T{d} \b{r}_2$, and the exact correlation energy functional vanishes
\begin{equation}
E_\T{c}[n_\delta,0]=0,
\label{HfcEc}
\end{equation}
where $n_\delta$ is the density for $\delta$ spin-up electron with $0\leq\delta\leq 1$. For range-separated DFT, the same conditions apply for systems with one or less spin-up electron, i.e. the exact short-range exchange energy functional cancels out the short-range Hartree energy
\begin{equation}
E_\T{x}^\T{sr}[n_\delta,0] = - E_\T{H}^\T{sr}[n_\delta],
\end{equation}
where the exact short-range exchange functional is defined as $E_\T{x}^\T{sr}[n_\uparrow,n_\downarrow]= -(1/2) \sum_\sigma \iint |\gamma_\sigma(\b{r}_1,\b{r}_2)|^2 w_\T{ee}^\T{sr}(|\b{r}_2-\b{r}_1|) \T{d} \b{r}_1 \T{d} \b{r}_2$, and the exact short-range correlation energy functional vanishes
\begin{equation}
E_\T{c}^\T{sr}[n_\delta,0]=0.
\end{equation}

\begin{figure*}
\includegraphics[width=.39\linewidth,clip=true,trim=0cm 0mm 0cm 0cm]{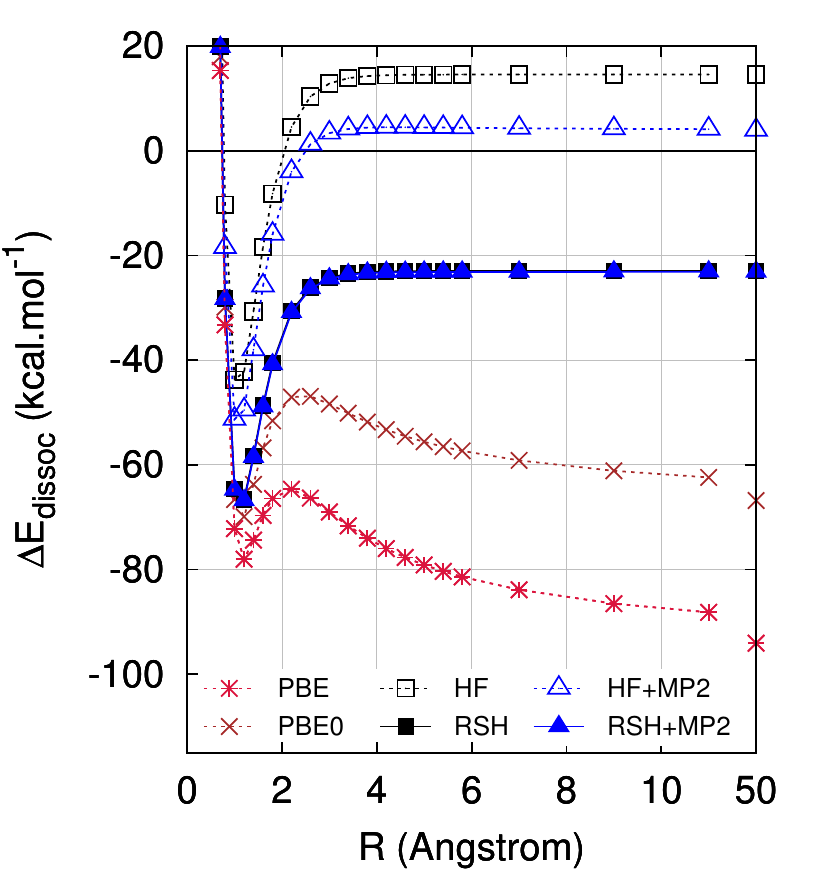}
\includegraphics[width=.39\linewidth,clip=true,trim=0cm 0mm 0cm 0cm]{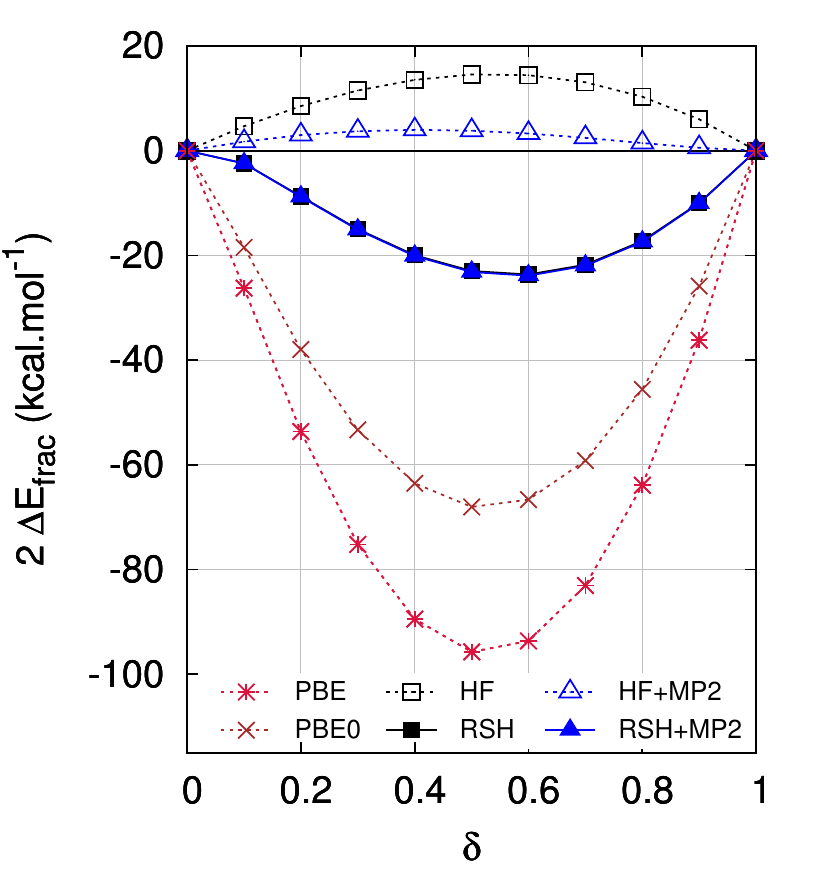}
\caption{Dissociation of He$_2^+$ and corresponding fractional-charge analysis for the He atom. On the left panel, the dissociation energy curve $\Delta E_\T{dissoc}(R)$ defined in Eq.~(\ref{he2+Edissoc}) is plotted as a function of the internuclear distance $R$ for each method. On the right panel, the fractional-charge error $2\; \Delta E_\T{frac}(\delta)$ defined in Eq.~(\ref{he2+Efrac}) is plotted as a function of $\delta$ for the same methods.}
\label{fig:he2+}
\end{figure*}

The deviations from these exact conditions for the fractional-charge H atom obtained with the full-range PBE exchange and correlation density functionals and with the short-range PBE exchange and correlation density functionals are plotted in Figure~\ref{fig:h_exc_frac}.
Clearly, except near $\delta=1$ (see below), the fractional-charge errors of both the PBE and RSH calculations are dominated by the error in the exchange functional.
While the full-range PBE and short-range PBE correlation functional give nearly identical correlation energies for the fractional-charge H atom for all $\delta$(which means that the LC and RSH schemes are nearly identical with this functional for this system), the short-range PBE exchange functional used in the RSH calculation has a fractional-charge error which is almost an order of magnitude smaller than that of the full-range PBE exchange functional for most $\delta$ (and it becomes comparable to the fractional-charge error of the correlation functional).
This is consistent with the fact that, in the limit of a very short-range electron-electron interaction (i.e., when the range-separation parameter $\mu$ is large), the exact short-range exchange density functional becomes a local functional of the density and therefore (semi)local-density approximations become exact~\cite{TouColSav-PRA-04}. In fact, a similar reduction of the fractional-charge errors in the H atom was already observed for the short-range LDA exchange-correlation density functional in Ref.~\onlinecite{Sav-CP-09}. Finally, note that the residual self-interaction error in the present RSH calculation for $\delta=1$ mainly comes from the short-range PBE correlation functional. As a matter of fact, the Kohn-Sham PBE calculation of the H atom (at $\delta=1$) largely benefits from a compensation of errors between the errors made by the PBE exchange and PBE correlation functionals. This compensation of errors is much reduced in the case of RSH. 

\subsection{Dissociation of He$_2^+$ and fractional-charge error in He}

The dissociation of He$_2^+$ allows one to discuss an example of the many-electron self-interaction error~\cite{MorCohYan-JCP-06,RuzPerCsoVydScu-JCP-06}. Similarly to the case of H$_2^+$, we show in Figure~\ref{fig:he2+} the dissociation curve calculated as
\begin{equation}
\Delta E_\T{dissoc}(R) = E(\T{He}_2^+,R) - \left[ E(\T{He}^+) + E(\T{He}) \right],
\label{he2+Edissoc}
\end{equation}
and the corresponding fractional-charge error
\begin{equation}
2\Delta E_\T{frac}(\delta) = 2E(\T{He}^{+(1-\delta)}) - 2\left[ (1-\delta) E(\T{He}^+) +\delta E(\T{He}) \right].
\label{he2+Efrac}
\end{equation}
Again, if spatial symmetry is not broken, we must have $\Delta E_\T{dissoc}(R\to\infty) = 2 \Delta E_\T{frac}(\delta=0.5)$, which can be seen by comparing the left and right panels of Figure~\ref{fig:he2+}.

Similar conclusions as in the case of H$_2^+$ can be drawn. PBE leads to large fractional-charge errors in the fractional He atom, and these errors are decreased a bit when using PBE0. Again, RSH shows much smaller fractional-charge errors. Just as for the fractional-charge H atom, the reduction of these errors essentially comes from the exchange part of the short-range PBE functional (not shown). Contrary to the case of H$_2^+$, here HF is not exact
and gives a concave curve as a function of $\delta$, with a positive fractional-charge error of about 15 kcal.mol$^{-1}$ at $\delta=0.5$. This indicates that, if spatial symmetry were allowed to break, the HF calculation of He$_2^+$ would give a lower energy curve with a dissociation limit corresponding to localized electrons on each fragment, He + He$^+$, and thus $\Delta E_\T{dissoc}(R\to\infty)$ would become zero (and the calculation would properly be size consistent). However, we are not interested in the case of symmetry breaking and we use the HF calculation of He$_2^+$ as displayed in Figure~\ref{fig:he2+} for subsequent MP2 calculations. Full-range HF+MP2 shows very small fractional-charge errors.
The range-separated RSH+MP2 method gives essentially the same fractional-charge errors as the errors obtained at the single-determinant RSH level. This is due to the fact that the long-range correlation energy is very small in this system.

\begin{figure*}
  \includegraphics[height=66mm,clip=true,trim=0mm 0mm 0cm 0cm]{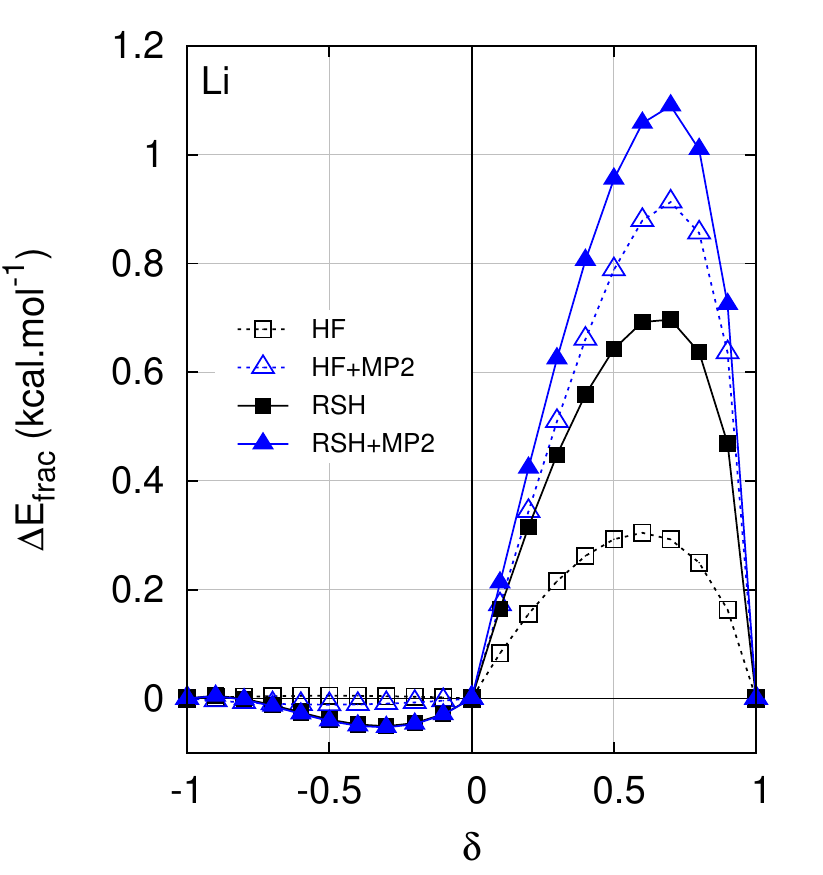}
  \includegraphics[height=66mm,clip=true,trim=7mm 0mm 0cm 0cm]{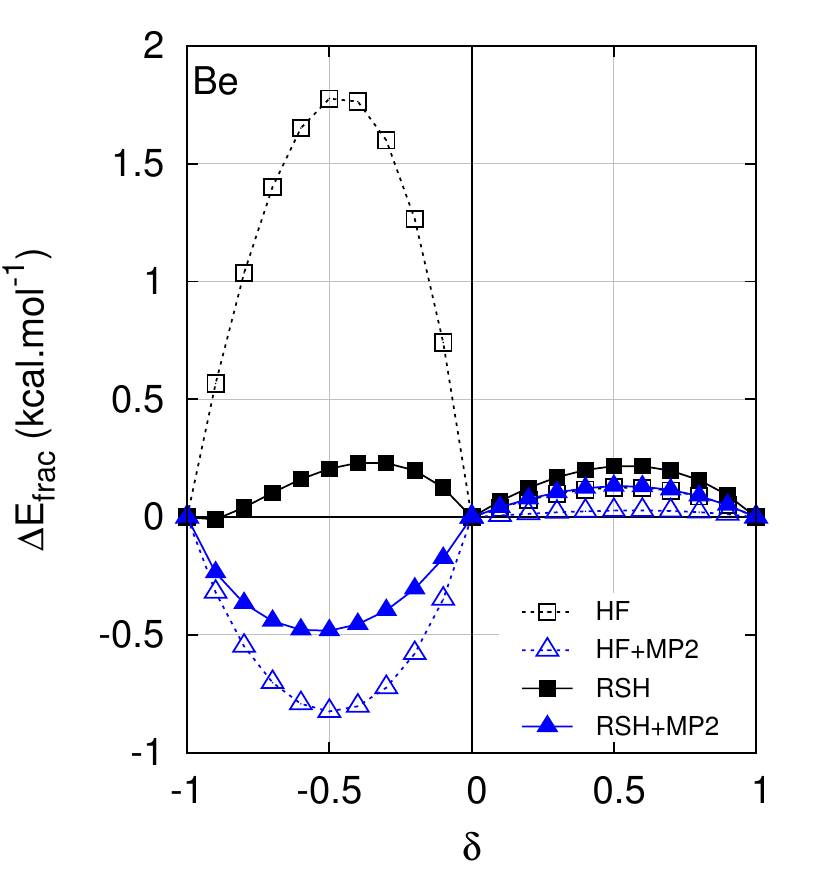}
  \includegraphics[height=66mm,clip=true,trim=7mm 0mm 0cm 0cm]{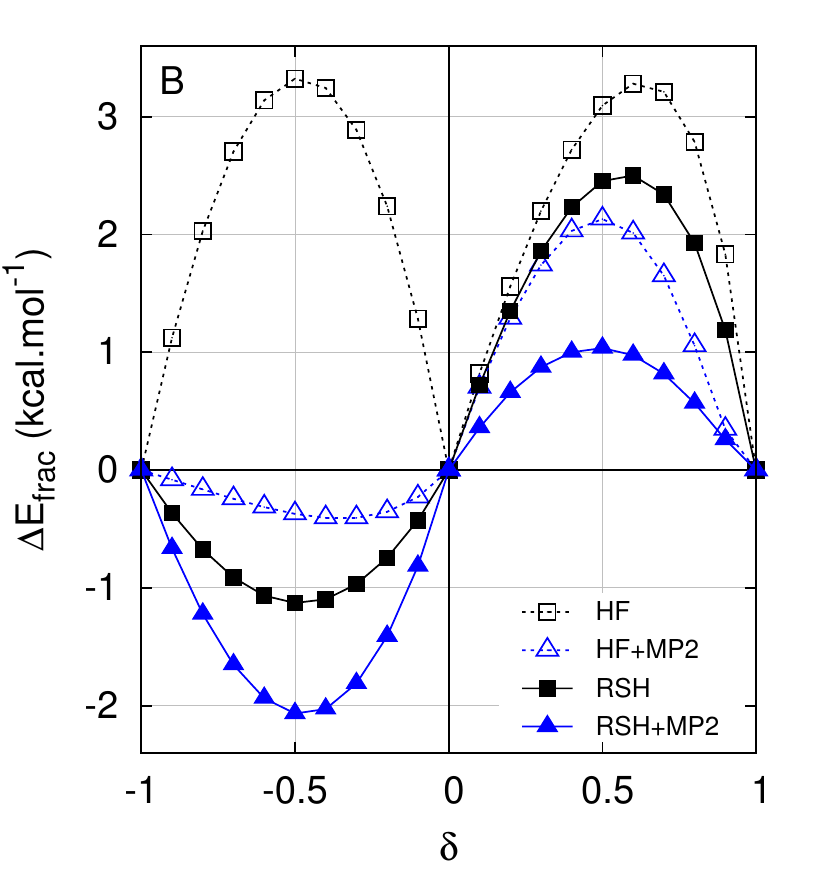}

  \includegraphics[height=66mm,clip=true,trim=0mm 0mm 0cm 0cm]{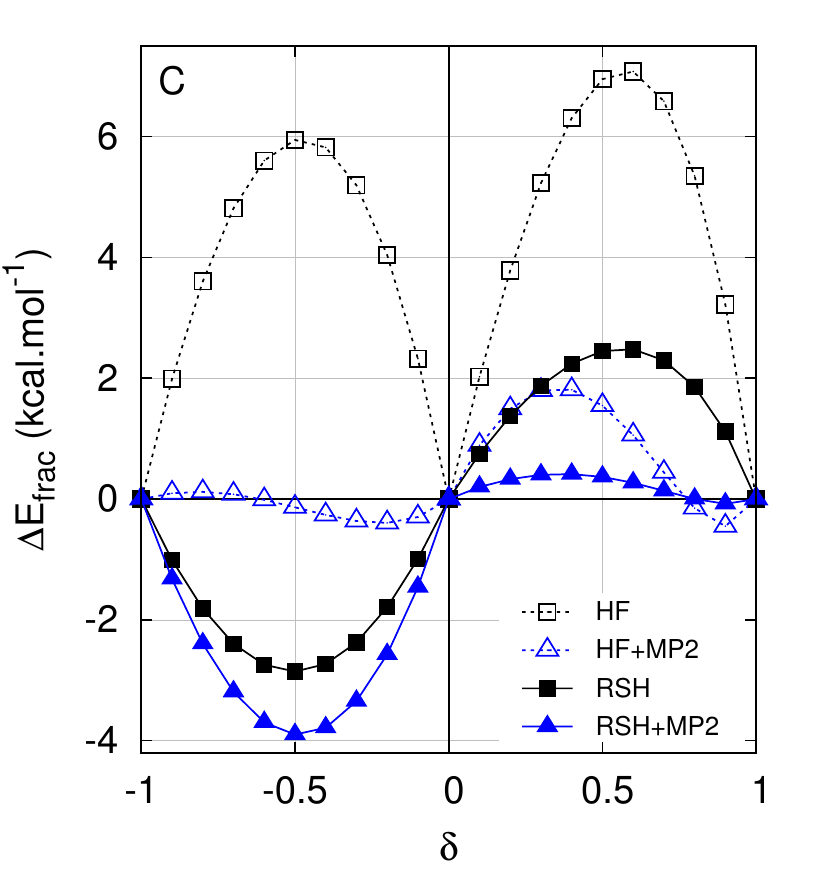}
  \includegraphics[height=66mm,clip=true,trim=7mm 0mm 0cm 0cm]{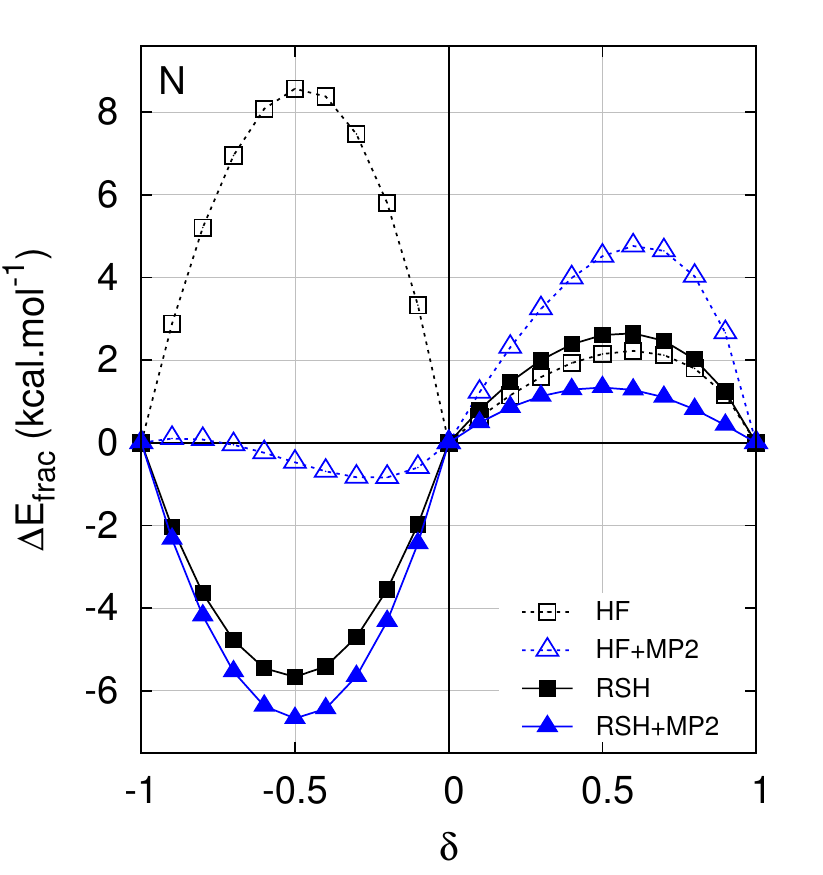}
  \includegraphics[height=66mm,clip=true,trim=7mm 0mm 0cm 0cm]{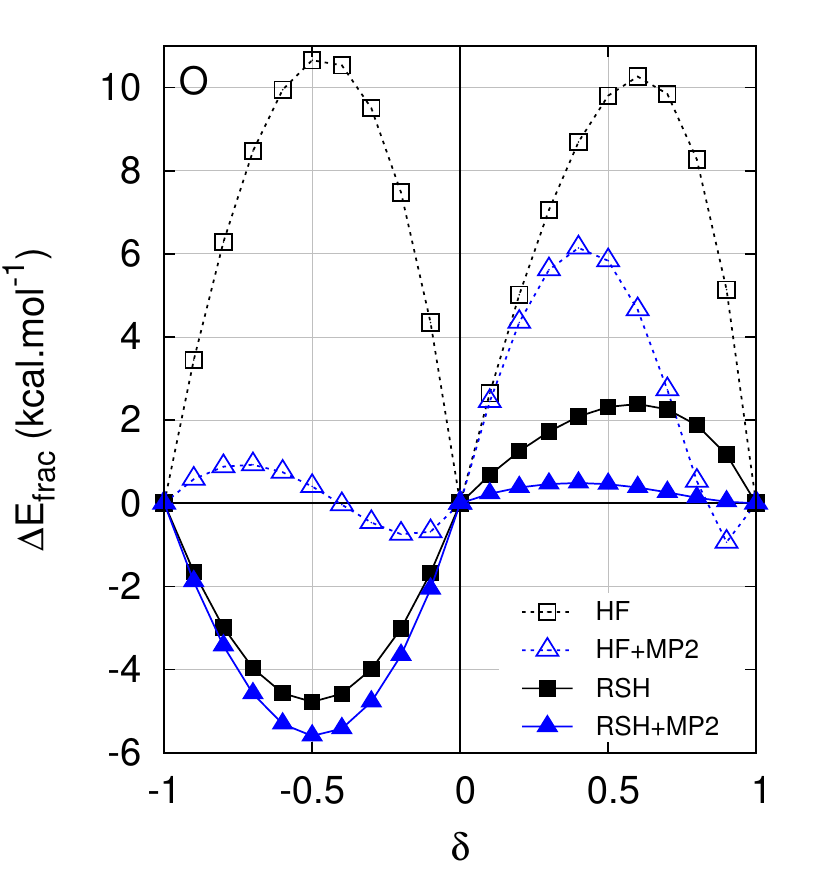}

  \includegraphics[height=66mm,clip=true,trim=0mm 0mm 0cm 0cm]{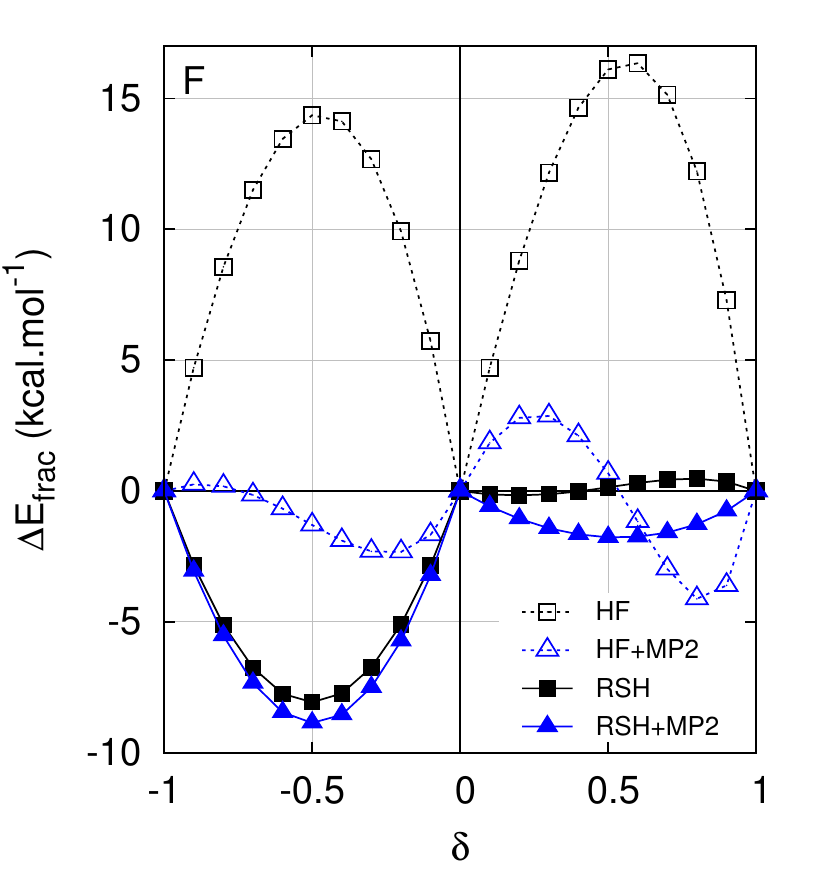}
  \includegraphics[height=66mm,clip=true,trim=7mm 0mm 0cm 0cm]{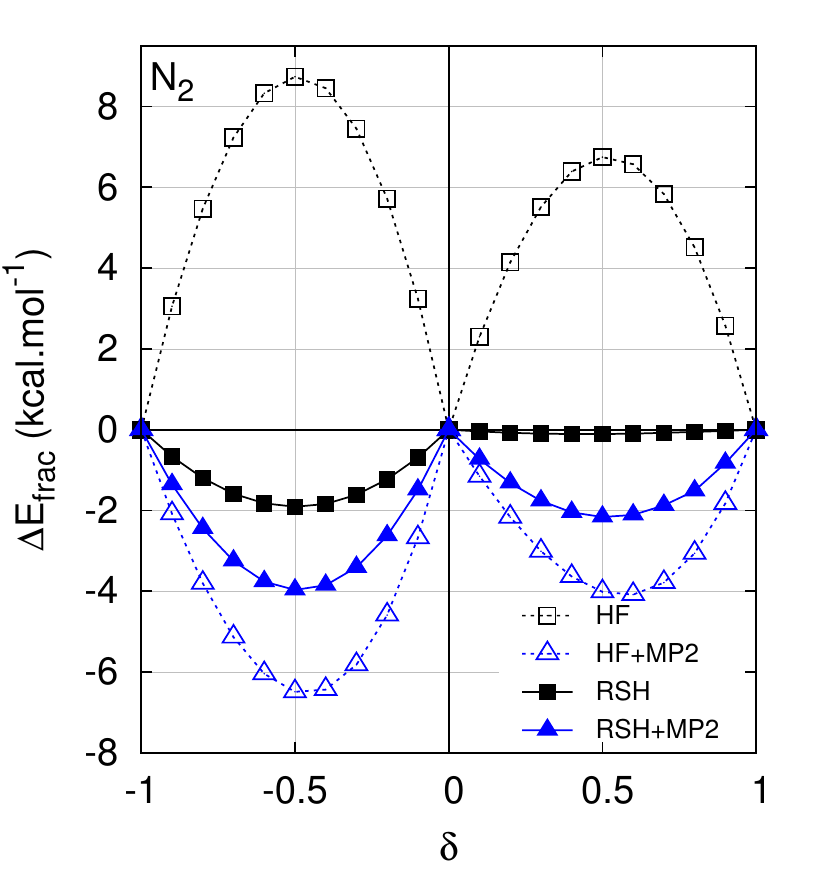}
  \includegraphics[height=66mm,clip=true,trim=7mm 0mm 0cm 0cm]{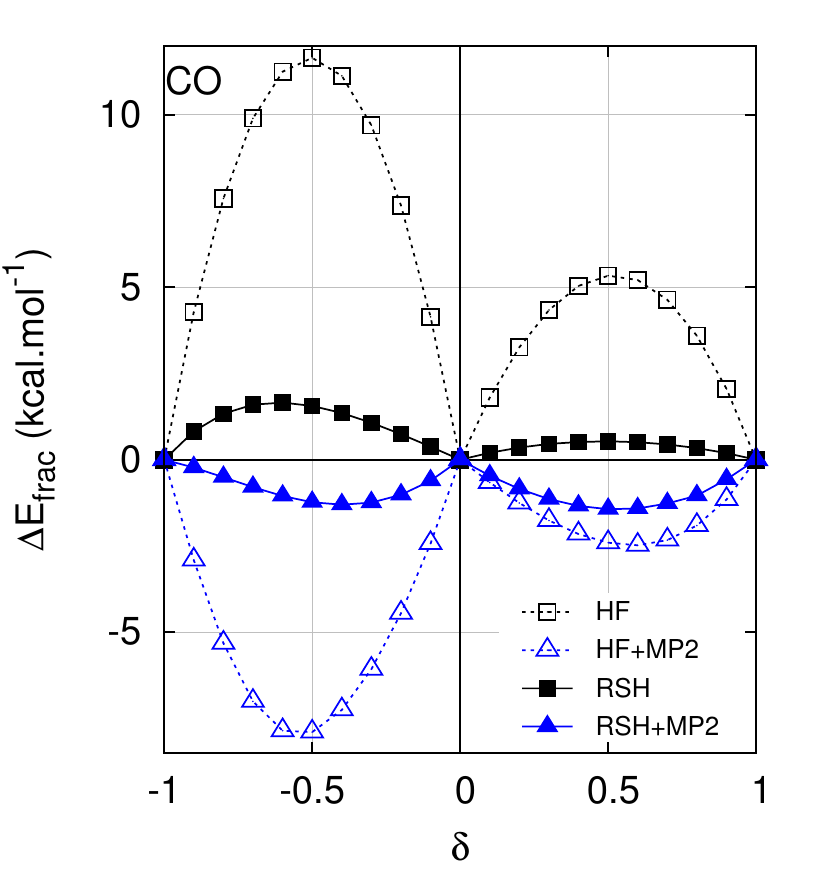}
  \caption{Fractional-charge errors of several methods for some atoms (Li, Be, B, C, N, O, F) and molecules (N$_2$, CO) with fractional numbers of electrons $Z=Z+\delta$, where $Z$ is the total charge of the nuclei and $-1\le\delta\le 1$. The errors $\Delta E_\T{frac}(\delta)$ defined in Eqs.~(\ref{eq:DeltaFrac+}) and (\ref{eq:DeltaFrac-}) are plotted as a function of $\delta$.}
  \label{fig:atom_anion}
\end{figure*}

\subsection{Fractional-charge errors in larger atoms (Li, Be, B, C, N, O, F) and in molecules (N$_2$, CO)}

We now investigate fractional-charge errors in larger atoms and in molecules. We consider the atoms Li, Be, B, C, N, O, F and the molecules N$_2$ and CO, and their fractional cations and anions with electron numbers $N=Z+\delta$ where $Z$ is the total charge of the nuclei and $-1\le\delta\le 1$. In Figure~\ref{fig:atom_anion}, we show the deviation from the exact piecewise linear behavior, i.e. 
\begin{equation}
\Delta E_\T{frac}(\delta) =
  E(\T{X}^{-\delta}) - \left[ -\delta E(\T{X}^+) +(1+\delta) E(\T{X}) \right],
\label{eq:DeltaFrac+}
\end{equation}
for $-1\le\delta\le 0$, and
\begin{equation}
\Delta E_\T{frac}(\delta) =
  E(\T{X}^{-\delta})   - \left[ (1-\delta) E(\T{X}) +\delta E(\T{X}^-) \right],
\label{eq:DeltaFrac-}
\end{equation}
for $0\le\delta\le 1$.

For the atomic systems, a first observation is that the maximal fractional-charge error tends to increase with $Z$. For almost all systems, HF shows a strong concave deviation. RSH gives fractional-charge errors that can be either convex or concave, and are generally smaller than in HF (the only exceptions being Li, and negatively charged Be and N). The behavior of the RSH fractional-charge errors can be understood as being intermediate between the concave fractional-charge errors of HF and the convex fractional-charge errors of Kohn-Sham PBE. The RSH fractional-charge errors are generally closer to the HF ones for negatively charged systems than for positively charged systems. As pointed in Ref.~\onlinecite{VydScuPer-JCP-07}, this can be rationalized by the fact that negatively charged systems are more diffuse than positively charged systems, and long-range HF exchange makes a larger contribution in such diffuse systems. In comparison to other methods, RSH gives particularly small errors for negatively charged F, N$_2$, and CO. The dependence of the accuracy of RSH on the size of the system suggests to tune the range-separation parameter in each system so as minimize the deviation from the exact piecewise linear behavior, as done in Refs.~\onlinecite{SteAutGovKroBae-JPCL-12,GlePeaToz-JCTC-13}. However, this approach has the disadvantage of being non size consistent~\cite{KarKroKum-JCP-13}, so we prefer to use a fixed value of the range-separation parameter, independent of the system. Finally, we mention that we have checked for the negatively charged F that the LC scheme (using the same short-range PBE exchange functional used in RSH) gives fractional-charge errors that are essentially identical to the RSH ones, meaning the long-range PBE correlation functional makes negligible contributions to the errors. We expect indeed that RSH and LC give very similar results in most systems.

Full-range MP2 almost always gives smaller fractional-charge errors than HF, with the exception of Li and negatively charged N. Depending on the system, the MP2 error curves are convex or concave, or have a S-shape as already observed in Refs.~\onlinecite{CohMorYan-JCTC-09,SuXu-JCTC-15}. The largest MP2 fractional-charge errors are obtained for the negatively charged atoms and for the molecules N$_2$ and CO. We note that the fractional-charge errors for negatively charged C reported in Ref.~\onlinecite{CohMorYan-JCTC-09} and for negatively charged O reported in Ref.~\onlinecite{SuXu-JCTC-15} appear to be significantly smaller than ours. We attribute these differences to the fact that we use basis sets augmented with diffuse basis functions, which has an important impact for negatively charged systems.

Range-separated MP2 gives fractional-charge errors than can be either smaller or larger than the RSH and full-range HF+MP2 ones, depending on the system. Specifically, in comparison to full-range HF+MP2, the RSH+MP2 fractional-charge errors are larger for the more compact systems (Li, Be, and positively charged B, C, N, O, F), and smaller for the more diffuse systems (negatively charged B, C, N, O, F, and the molecules N$_2$ and CO).

\begin{figure*}
\includegraphics[width=.38\linewidth,clip=true,trim=0cm 0mm 0cm 0cm]{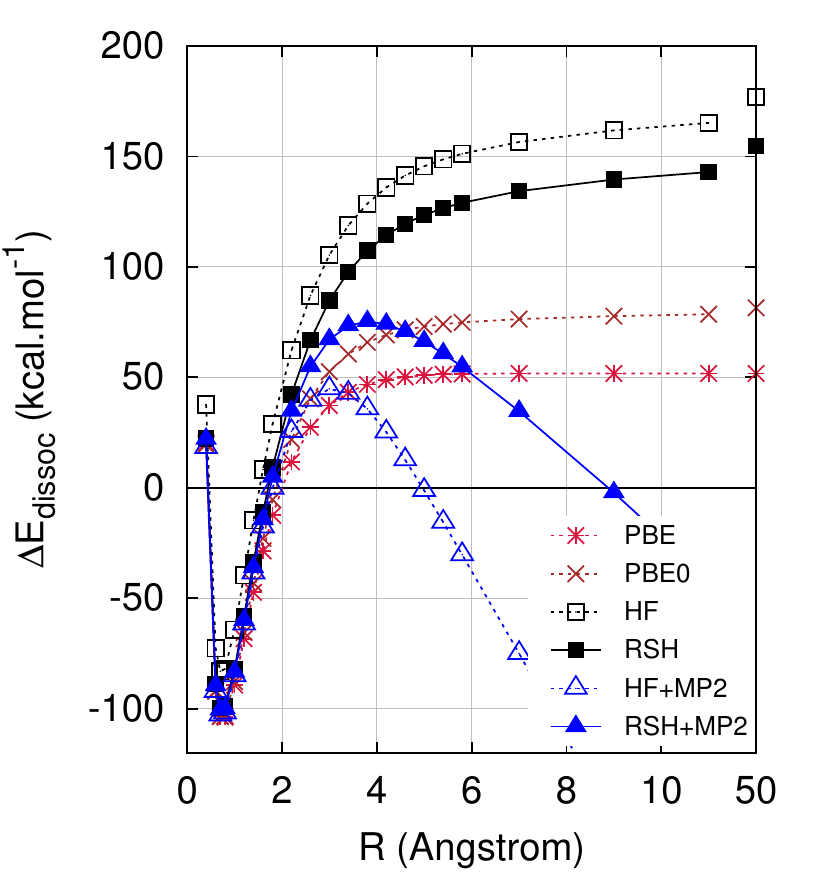}
\includegraphics[width=.38\linewidth,clip=true,trim=0cm 0mm 0cm 0cm]{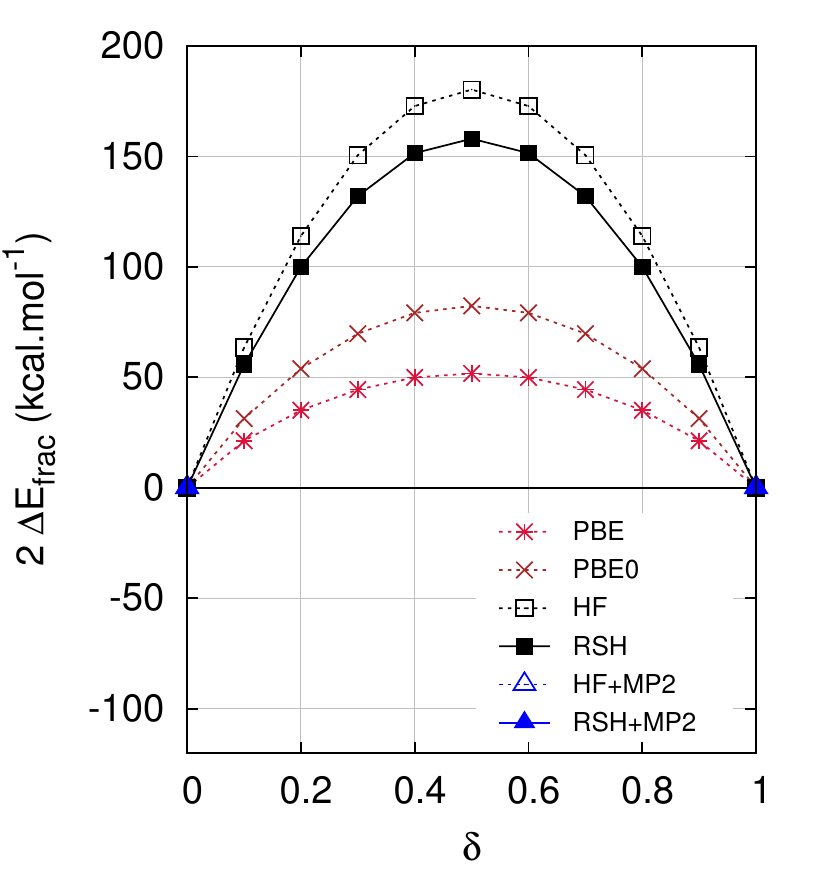}
\caption{Dissociation of H$_2$ and corresponding fractional-spin analysis for the H atom. On the left panel, the dissociation energy curve $\Delta E_\T{dissoc}(R)$ defined in Eq.~(\ref{h2Edissoc}) is plotted as a function of the internuclear distance $R$ for each method. On the right panel, the fractional-spin error $2\Delta E_\T{frac}(\delta)$ defined in Eq.~(\ref{h2Efrac}) is plotted as a function of $\delta$ for the same methods. The HF+MP2 and RSH+MP2 energies of H$_2$ diverge to $-\infty$ in the dissociation limit and their corresponding fractional-spin errors in the H atom also diverge for all $0 <\delta < 1$.}
\label{fig:h2}
\end{figure*}

\subsection{Dissociation of H$_2$ and fractional-spin error in H}

The dissociation of H$_2$ is a prototypical example for studying the static-correlation error. In the left panel of Figure~\ref{fig:h2}, the total energy of H$_2$ calculated by each method, preserving spin symmetry in the calculation, is plotted as a function of the internuclear distance $R$, taking as zero energy reference twice the energy of the isolated atom $\T{H}$ calculated with the same method
\begin{equation}
\Delta E_\T{dissoc}(R) = E(\T{H}_2,R) - 2 E(\T{H}).
\label{h2Edissoc}
\end{equation}
On the right panel of Figure~\ref{fig:h2}, for each corresponding method, we plot twice the error in the total energy of a H atom with fractional spin-up and spin-down electron numbers $N_\uparrow = 1-\delta$ and $N_\downarrow = \delta$, denoted as $\T{H}^{1-\delta,\delta}$, with respect to the energy of the normal H atom as a function of $\delta$ with $0\leq\delta\leq 1$
\begin{equation}
2\Delta E_\T{frac}(\delta) = 2 E(\T{H}^{1-\delta,\delta}) - 2 E(\T{H}).
\label{h2Efrac}
\end{equation}
According to the constancy condition of Eq.~(\ref{ENupdn}), we should have $\Delta E_\T{frac}(\delta)=0$ for all $0\leq\delta\leq 1$. For all the SCF methods considered here, H$_2$ dissociates into two hydrogen atoms with each half spin-up and half spin-down electron, i.e. $E(\T{H}_2,R\to\infty) = E(\T{H}^{0.5,0.5} + \T{H}^{0.5,0.5}) = 2 E(\T{H}^{0.5,0.5})$. The dissociation limit of each energy curve is thus
\begin{eqnarray}
  \label{eq:h2dissocfrac}
\Delta E_\T{dissoc}(R\to\infty) &=& 2 E(\T{H}^{0.5,0.5}) - 2 E(\T{H})
\nonumber\\
&=& 2\Delta E_\T{frac}(\delta=0.5).
\end{eqnarray}
Thus, the value $2\Delta E_\T{frac}(\delta=0.5)$ corresponds to the static-correlation error in H$_2$ in the dissociation limit. More generally, the curve of the fractional-spin error $\Delta E_\T{frac}(\delta)$ is a convenient way to analyze the static-correlation error of a method.

As well known, (spin-restricted) HF gives a large static-correlation error in H$_2$ in the dissociation limit, or equivalently a large fractional-spin error in the H atom at $\delta=0.5$ (about 180 kcal.mol$^{-1}$). RSH only decreases a bit the fractional-spin errors (about 160 kcal.mol$^{-1}$ at $\delta=0.5$). PBE0 has much less fractional-spin errors (about 80 kcal.mol$^{-1}$ at $\delta=0.5$). Expectedly, the PBE calculation, which does not include any HF exchange, leads to the smallest fractional-spin errors (about 50 kcal.mol$^{-1}$ at $\delta=0.5$). The HF+MP2 and RSH+MP2 energies of H$_2$ diverge to $-\infty$ in the dissociation limit. Their corresponding fractional-spin errors in the H atom also diverge for all $0 <\delta < 1$ due to the contribution from $i=a$ and $j=b$ in Eq.~(\ref{Td1}) where $i$ and $j$ are the degenerate fractionally occupied spin-up and spin-down orbitals. In fact, as noted in Ref.~\onlinecite{CohMorYan-JCTC-09}, the MP2 correlation energy always diverges if more than one spin orbital within a degenerate set is fractionally occupied.

As for the fractional-charge H atom, it is interesting to analyze the contributions to the fractional-spin error coming from the approximate exchange and correlation density functionals used. In Kohn-Sham DFT, since the fractional-spin H atom contains only one electron, the exact exchange-correlation functional cancels out the Hartree energy
\begin{equation}
E_\T{xc}[n_{1-\delta},n_{\delta}]=- E_\T{H}[n_{1-\delta}+n_{\delta}],
\label{ExcHspin}
\end{equation}
when $n_{1-\delta}$ and $n_{\delta}$ are the spin-up and spin-down densities, respectively. However, the correlation part does not vanish, contrary to the case of the fractional-charge H atom [see Eq.~(\ref{HfcEc})]. This can be seen as follows. The Hartree energy can be decomposed as
\begin{equation}
E_\T{H}[n_{1-\delta}+n_{\delta}]=E_\T{H}[n_{1-\delta}] + E_\T{H}[n_{\delta}] + U[n_{1-\delta},n_{\delta}],
\end{equation}
where $U[n_{1-\delta},n_{\delta}]$ is the Coulomb interaction between the spin-up and spin-down densities
\begin{equation}
U[n_{1-\delta},n_{\delta}] =\iint n_{1-\delta}(\b{r}_1) n_{\delta}(\b{r}_2) w_\T{ee}(|\b{r}_2-\b{r}_1|) \T{d} \b{r}_1 \T{d} \b{r}_2.
\end{equation}
Since the exact exchange functional cancels out only the Hartree energy of each separate spin density
\begin{equation}
E_\T{x}[n_{1-\delta},n_{\delta}]=- \left( E_\T{H}[n_{1-\delta}] + E_\T{H}[n_{\delta}] \right),
\label{ExHspin}
\end{equation}
the exact correlation energy functional must cancel the term $U[n_{1-\delta},n_{\delta}]$
\begin{equation}
E_\T{c}[n_{1-\delta},n_{\delta}]=- U[n_{1-\delta},n_{\delta}].
\label{HfsEc}
\end{equation}
The correlation energy in Eq.~(\ref{HfsEc}) is clearly characteristic of static (or strong) correlation: it is first order in the electron-electron interaction and it has only a correlation potential contribution (and not a correlation kinetic contribution). The term $U[n_{1-\delta},n_{\delta}]$ is a spurious interaction between the densities of the two elements of the spin-up/down ensemble. As pointed out in Ref.~\onlinecite{GouDob-JCP-13}, it is analogous to the unphysical ``ghost'' interaction of ensemble DFT for excited states~\cite{GidPapGro-PRL-02} (see, also, Refs.~\onlinecite{Sav-INC-96,Sav-CP-09}). This term could be removed by an ad hoc correction~\cite{MorCohYan-PRL-09} or, more generally, by a redefinition of the Hartree/exchange/correlation decomposition for situations with fractional-occupation numbers~\cite{GouDob-JCP-13}. For range-separated DFT, similar expressions apply for the fractional-spin H atom. The exact short-range exchange functional is
\begin{equation}
E_\T{x}^\T{sr}[n_{1-\delta},n_{\delta}]=- \left( E_\T{H}^\T{sr}[n_{1-\delta}] + E_\T{H}^\T{sr}[n_{\delta}] \right),
\label{ExsrHspin}
\end{equation}
and the exact short-range correlation functional is
\begin{equation}
E_\T{c}^\T{sr}[n_{1-\delta},n_{\delta}]=- U^\T{sr}[n_{1-\delta},n_{\delta}],
\label{HfsEcsr}
\end{equation}
where $U^\T{sr}[n_{1-\delta},n_{\delta}]$ is a spurious short-range interaction between the spin-up and spin-down densities
\begin{equation}
U^\T{sr}[n_{1-\delta},n_{\delta}] =\iint n_{1-\delta}(\b{r}_1) n_{\delta}(\b{r}_2) w_\T{ee}^\T{sr}(|\b{r}_2-\b{r}_1|) \T{d} \b{r}_1 \T{d} \b{r}_2.
\end{equation}
Again, this spurious ``ghost'' interaction $U^\T{sr}[n_{1-\delta},n_{\delta}]$ could be removed from the correlation term by a redefinition of the short-range Hartree/exchange/correlation decomposition in the ensemble formalism (see in particular Ref.~\onlinecite{SenHedAlaKneFro-MP-16}).

\begin{figure}
\includegraphics[width=1\linewidth,clip=true,trim=0cm 0mm 0cm 0cm]{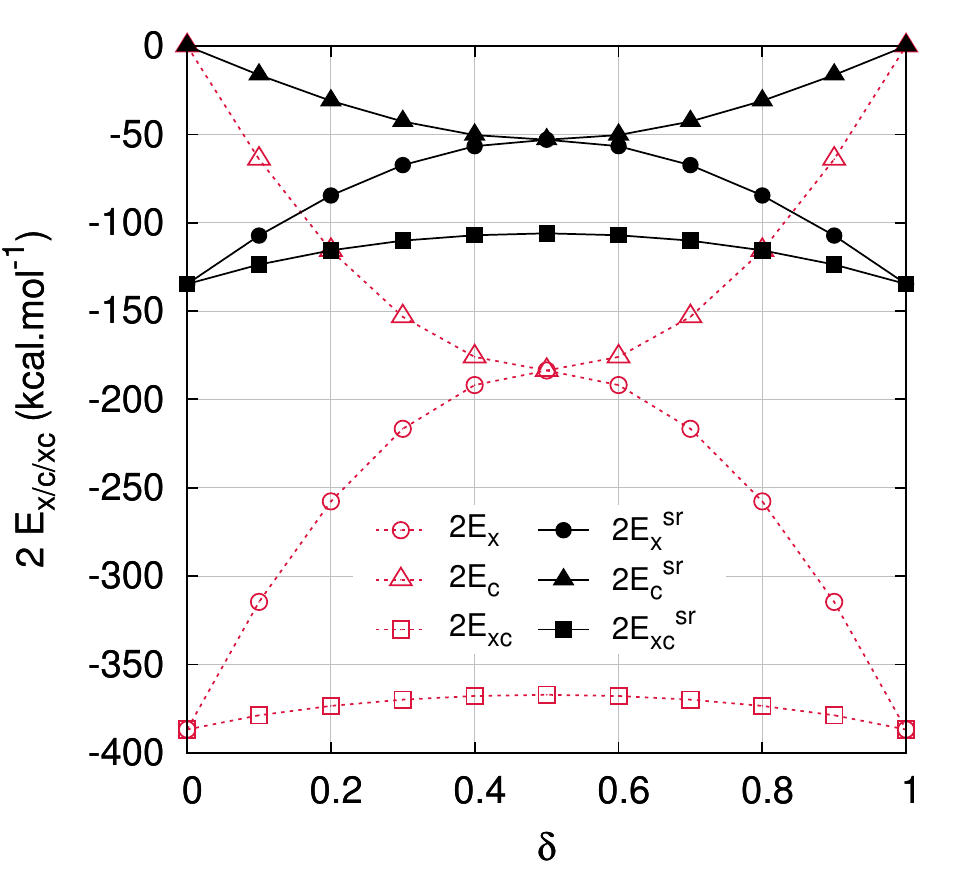}
\caption{``Exact'' exchange and correlation energies in the H atom with fractional spin-up and spin-down electron numbers $N_\uparrow = 1-\delta$ and $N_\downarrow = \delta$ as a function of $\delta$. For the full-range case, the exchange and correlation energies were calculated as $E_\T{x}[n_{1-\delta},n_{\delta}]=- \left( E_\T{H}[n_{1-\delta}] + E_\T{H}[n_{\delta}] \right)$ and $E_\T{c}[n_{1-\delta},n_{\delta}]=- U[n_{1-\delta},n_{\delta}]$, where the spin densities $n_{1-\delta}$ and $n_\delta$ are not exact but obtained from a self-consistent PBE calculation at the value of $\delta$. For the range-separated case, the short-range exchange and correlation energies were calculated as $E_\T{x}^\T{sr}[n_{1-\delta},n_{\delta}]=- \left( E_\T{H}^\T{sr}[n_{1-\delta}] + E_\T{H}^\T{sr}[n_{\delta}] \right)$ and $E_\T{c}^\T{sr}[n_{1-\delta},n_{\delta}]=- U^\T{sr}[n_{1-\delta},n_{\delta}]$, where the spin densities $n_{1-\delta}$ and $n_\delta$ are not exact but obtained from a self-consistent RSH calculation at the value of $\delta$. The energies are multiplied by a factor of 2 to be directly comparable to Figure~\ref{fig:h2}.}
\label{fig:h_spin_exc_exact}
\end{figure}

\begin{figure}
\includegraphics[width=1\linewidth,clip=true,trim=0cm 0mm 0cm 0cm]{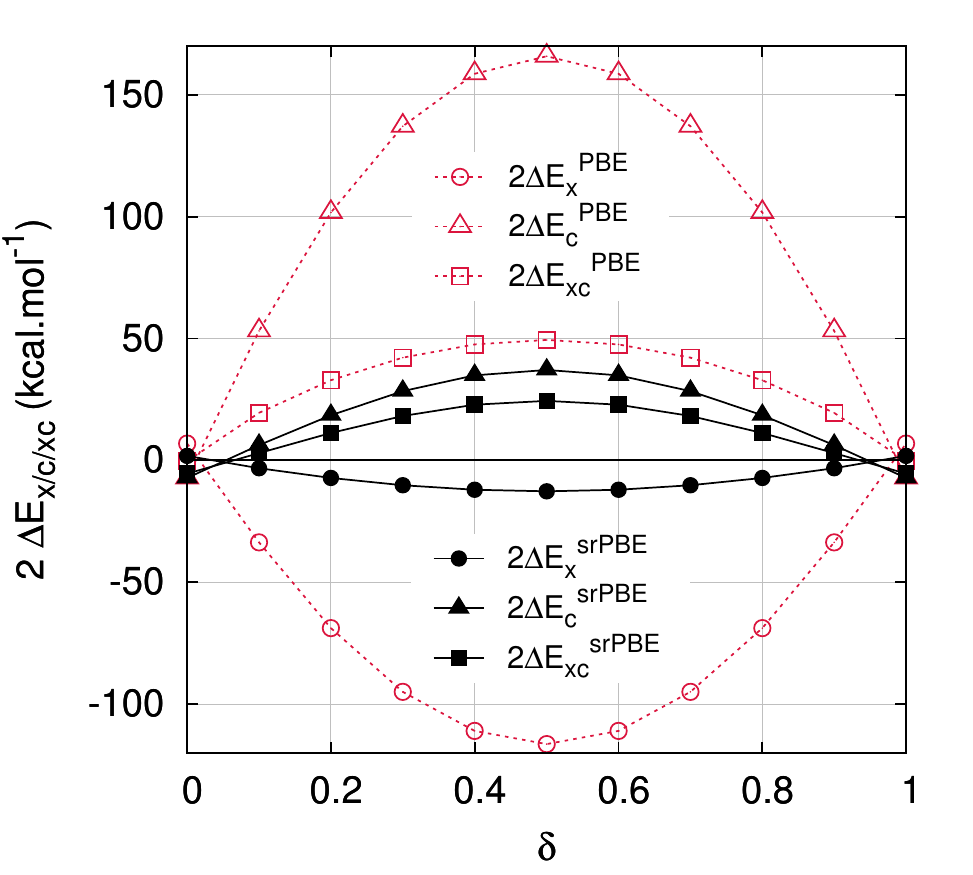}
\caption{Density-functional-approximation contributions to the fractional-spin error in the H atom with fractional spin-up and spin-down electron numbers $N_\uparrow = 1-\delta$ and $N_\downarrow = \delta$ as a function of $\delta$. For Kohn-Sham PBE calculations, we show the error due to the PBE exchange energy, $\Delta E_\T{x}^\T{PBE} = E_\T{x}^\T{PBE}[n_{1-\delta},n_\delta] + E_\T{H}[n_{1-\delta}] + E_\T{H}[n_{\delta}]$, the PBE correlation energy $\Delta E_\T{c}^\T{PBE} = E_\T{c}^\T{PBE}[n_{1-\delta},n_\delta] + U[n_{1-\delta},n_\delta]$, and their sum $\Delta E_\T{xc}^\T{PBE}=\Delta E_\T{x}^\T{PBE}+\Delta E_\T{c}^\T{PBE}$, where the spin densities $n_{1-\delta}$ and $n_\delta$ are from the self-consistent PBE calculation at the value of $\delta$. For the RSH calculations, we show the error due to the short-range PBE exchange energy, $\Delta E_\T{x}^\T{srPBE} = E_\T{x}^\T{srPBE}[n_{1-\delta},n_\delta] + E_\T{H}^\T{sr}[n_{1-\delta}] + E_\T{H}^\T{sr}[n_{\delta}]$, the short-range PBE correlation energy $\Delta E_\T{c}^\T{srPBE} = E_\T{c}^\T{srPBE}[n_{1-\delta},n_\delta] + U^\T{sr}[n_{1-\delta},n_\delta]$, and their sum $\Delta E_\T{xc}^\T{srPBE} = \Delta E_\T{x}^\T{srPBE} + \Delta E_\T{c}^\T{srPBE}$, where the spin densities $n_{1-\delta}$ and $n_\delta$ are from the self-consistent RSH calculation at the value of $\delta$. The errors are multiplied by a factor of 2 to be directly comparable to Figure~\ref{fig:h2}.
}
\label{fig:h_spin_pbe_rsh_exc}
\end{figure}

In Figure~\ref{fig:h_spin_exc_exact}, we show as a function of $\delta$ the full-range exchange and correlation energies of Eqs.~(\ref{ExHspin}) and (\ref{HfsEc}), and the short-range exchange and correlation energies of Eqs.~(\ref{ExsrHspin}) and (\ref{HfsEcsr}). In principle, in this system, the full-range exchange-correlation energy of Eq.~(\ref{ExcHspin}), or its short-range variant, is constant with respect to $\delta$ since in the exact theory $n_{1-\delta} + n_{\delta}=n$ where $n$ is the density of the H atom independent of $\delta$. However, due to the fact that we have used approximate spin densities $n_{1-\delta}$ and $n_{\delta}$ obtained from a self-consistent PBE or RSH calculation at each value of $\delta$, in practice the full-range or short-range exchange-correlation energies are only approximately constant with respect to $\delta$. For both the full-range and short-range cases, the correlation contribution is zero at $\delta=0$ and $\delta=1$, and passes by a maximum (in absolute value) at $\delta=0.5$, where the exchange and correlation energies are equal. The exchange contribution exhibits a symmetrically opposite behavior (being minimum in absolute value at $\delta=0.5$) so as to make the sum of the exchange and correlation energies approximately constant with respect to $\delta$. This illustrates the fact that the decomposition into exchange and correlation is somewhat artificial for strongly correlated systems. The large HF fractional-spin error reported in Figure~\ref{fig:h2} obviously comes from the fact that in HF only the exchange contribution is included and the equally important correlation contribution is neglected. The short-range exchange and correlation energies are about four times smaller than the full-range exchange and correlation energies, but they behave similarly with respect to $\delta$.

The deviations from the exact expressions of Eqs.~(\ref{ExHspin}),~(\ref{HfsEc}),~(\ref{ExsrHspin}),~(\ref{HfsEcsr}) obtained with the PBE exchange and correlation density functionals and with the short-range PBE exchange and correlation density functionals are plotted in Figure~\ref{fig:h_spin_pbe_rsh_exc}. The fractional-spin errors stemming from the full-range PBE exchange and full-range PBE correlation functionals are large but of opposite signs, leading to a large compensation of errors between exchange and correlation
and thus to the relatively good performance of Kohn-Sham PBE for the dissociation of H$_2$. Both the short-range PBE exchange and short-range PBE correlation functionals have considerably smaller fractional-spin errors than their full-range counterparts. We note that, as for the fractional-charge H atom, the full-range PBE correlation energy and short-range PBE correlation energy are nearly identical (not shown) in this system, so the smaller errors in the short-range PBE correlation functional are in fact due to the smaller magnitude of the exact short-range correlation energy $-U^\T{sr}[n_{1-\delta},n_{\delta}]$ compared to the exact full-range correlation energy $-U[n_{1-\delta},n_{\delta}]$ (as shown in Figure~\ref{fig:h_spin_exc_exact}). However, the compensation of errors between short-range exchange and short-range correlation is less dramatic than in the full-range case, and the fractional-spin errors in the short-range PBE exchange-correlation functional are reduced by only about of factor of 2 compared to the full-range PBE exchange-correlation functional. At first sight, the fact that the fractional-spin errors in the short-range PBE functional are smaller than the ones of the full-range PBE functional seems to be in contradiction with the worse performance of RSH compared to Kohn-Sham PBE for the fractional-spin H atom seen in Figure~\ref{fig:h2}. In fact, the additional error seen in the RSH results comes from the fact that the long-range HF exchange energy is not compensated by an appropriate long-range correlation energy. If the long-range correlation energy were calculated by methods capable of dealing with static correlation, such as multiconfiguration self-consistent field (MCSCF) or density-matrix functional theory (DMFT), then the fractional-spin errors would only come from the short-range exchange-correlation density functional and would be smaller than the errors obtained in Kohn-Sham calculations. It is indeed what has been observed in range-separated MCSCF+DFT~\cite{PedJen-JJJ-XX,FroTouJen-JCP-07,ShaSavJenTou-JCP-12} and range-separated DMFT+DFT~\cite{RohTouPer-PRA-10,RohPer-JCP-11} calculations of H$_2$ in the dissociation limit.

\section{Conclusion}

In this work, we have investigated fractional-charge and fractional-spin errors in atoms and molecules obtained with range-separated DFT schemes, namely the RSH method which combines long-range HF exchange with a short-range PBE exchange-correlation functional, and the RSH+MP2 method which adds long-range MP2 correlation. Very similarly to the LC scheme, RSH gives much smaller fractional-charge errors than standard Kohn-Sham applied with the semilocal PBE or hybrid PBE0 approximation. RSH also generally leads to smaller fractional-charge errors than standard HF. As regards RSH+MP2, it tends to have smaller fractional-charge errors than standard MP2 for the most diffuse systems (molecules and negatively charged atoms) but larger fractional-charge errors for the more compact systems (positively charged atoms).

Even though the individual contributions to the fractional-spin errors in the H atom coming from the short-range PBE exchange and correlation density functionals are smaller than the corresponding contributions for the full-range PBE exchange and correlation density functionals, RSH gives fractional-spin errors that are larger than in standard Kohn-Sham PBE and only slightly smaller than in standard HF. Moreover, adding long-range MP2 correlation only leads to infinite fractional-spin errors. This points to the necessity of accounting for long-range static correlation by appropriate methods, e.g. MCSCF, DMFT, or certain variants of RPA. Only with such approaches, together with improved short-range exchange-correlation approximations, one can expect to have range-separated schemes with both small fractional-charge and fractional-spin errors. Work in this direction is underway.

\section*{Acknowledgements}
It is a great pleasure to dedicate this paper to Hans J{\o}rgen Aagaard Jensen who made outstanding contributions to computational quantum chemistry and in particular to the development of range-separated DFT. Contract grant sponsors: Labex CalSimLab part of French funds managed by the ANR within the Investissements d'Avenir programme; contract grant numbers: ANR-11-LABX-0037-01.


\end{document}